# Observation of Temporal Reflection and Broadband Frequency Translation at Photonic Time-Interfaces

Hady Moussa[1,2]*, Gengyu Xu[1]*, Shixiong Yin[1,2]*, Emanuele Galiffi[1], Younes Ra'di[1,3] and Andrea Alù[1,2,4]

[1] Photonics Initiative, Advanced Science Research Center, City University of New York; New York, NY 10031, United States

[2] Department of Electric Engineering, City College of City University of New York; New York, NY 10031, United States

[3] Department of Electrical Engineering and Computer Science, Syracuse University, Syracuse, NY 13244, United States

[4] Physics Program, Graduate Center, City University of New York; New York, NY 10016, United States

Corresponding author. Email: aalu@gc.cuny.edu

* These authors contributed equally to the work.

**Time-reflection is a uniform inversion of the temporal evolution of a signal, which arises when an abrupt change in the properties of the host material occurs uniformly in space.[1-3] At such a time-interface, a portion of the input signal is time-reversed, and its frequency spectrum is homogeneously translated while its momentum is conserved, forming the temporal counterpart of a spatial interface. Combinations of time-interfaces, forming time-metamaterials and Floquet matter[4], exploit the interference of multiple time-reflections for extreme wave manipulation, leveraging time as a new degree of freedom[5]. Here, we report the observation of photonic time-reflection and associated broadband frequency translation in a switched transmission-line metamaterial whose effective capacitance is homogeneously and abruptly changed via a synchronized array of switches. A pair of temporal interfaces are combined to demonstrate time-reflection-induced wave interference, realizing the temporal counterpart of a Fabry-Perot cavity[6]. Our results establish the foundational building blocks to realize time-metamaterials and Floquet photonic crystals, with opportunities for extreme photon manipulation in space and time[7-8].**

Reflection is a universal phenomenon occurring when a traveling wave encounters an inhomogeneity. Spatial reflections arise at a sharp discontinuity in space: here, momentum is exchanged between an incoming wave and the interface, which acts as a momentum bath, while frequency is conserved. As the basis of wave scattering, spatial reflections play a key role in wave control and routing, as well as in the formation of resonant modes, filtering, band engineering and metamaterial responses. Recently, advances across nonlinear wave sciences have stirred significant interest in the use of time as a new degree of freedom for wave scattering, leveraging time-varying media as reservoirs that mix and exchange energy with the waves in the system. As examples of these opportunities, photonic time-crystals and Floquet wave phenomena have raised interest across the broader physics community[9-16]. In this context, time-reflection (TR) constitutes the temporal counterpart of spatial reflection, with dual features. This effect occurs at a time-interface, i.e., when the properties of the host medium are switched homogeneously in space over a timespan much faster than the wave dynamics. Upon TR, an input wave is partly time-reversed: its energy and frequency content are generally transformed, while momentum is conserved because of spatial translational symmetry.

Time-reversal is a key functionality for a variety of applications, from channel estimation in communication systems, to compensation of signal distortion and dispersion. The most common way of realizing time-reversal is through the digitization and retransmission of a recorded signal through a computer[17], but with significant requirements in terms of processing time and energy, as well as memory demands. In the analog domain, time-reversal can be achieved by periodically modulating the properties of the host medium at twice the frequency of the signal. This phenomenon has been observed in acoustics[18], in magnonics[19] and, for electromagnetic waves, both at radio-frequencies[17] and, with lower efficiency, in optics[20]. However, parametric phenomena are inherently slow and narrowband, relying on extended exposure of the signal of interest to a periodic modulation driving the resonant coupling between positive and negative frequencies[20], and hence subject to instabilities leading to highly dispersive and nonlinear distortions. On the contrary, TR at a time-interface enables ultrafast, and ultrabroadband time-reversal and, where desirable, efficient frequency translation of an arbitrary waveform. While several exciting theoretical proposals have been put forward to exploit these features for a variety of exotic photonic functionalities, including subwavelength focusing[21] imaging through random media[22], temporal anti-reflection coatings and advanced frequency filtering[6], inverse prisms[23], temporal aiming[24], analog computing[25], and the ultrafast generation of squeezed states of quantum light[26], time-interfaces have so far only been observed for water waves[3], remaining elusive to photonics and thus drastically limiting their impact. The key challenge in this quest consists in designing and realizing a setup capable of imparting sufficiently strong and fast variations to the electromagnetic properties of a material uniformly in space, hence requiring a metamaterial featuring a temporal response much faster than the temporal wave dynamics. Energy and dispersion requirements may also become very demanding when imparting such strong and fast modulations of the material properties[27], factors that have been hindering the experimental demonstration of TR in electromagnetics to date.

Here, we are able to tackle these challenges, and demonstrate time-interfaces and TR in a microwave transmission-line metamaterial (TLM) periodically loaded by a deeply subwavelength array of lumped capacitors, synchronously added to, or removed from, a microstrip through voltage-controlled reflective switches. Upon switching, we can uniformly and strongly change the effective capacitance per unit length of the TLM much faster than the temporal variations of the broadband signals propagating through it. This realizes a time-interface, with associated photonic TR, as well as broadband, efficient frequency translation, as conceptually shown in Fig. 1a. As we discuss in the following, by switching in and out the capacitive loads we are able to modify drastically and abruptly the electromagnetic properties of the TLM without affecting its linear dispersion and without large energy requirements. By implementing a pair of such time-interfaces, we form a temporal slab in which the reflected and refracted signals at each time-interface interfere, demonstrating the temporal analogue of a Fabry-Perot filter[6]. Our results establish the fundamental building blocks to exploit time as a new degree of freedom for extreme wave manipulation in metamaterials[4,5,28].

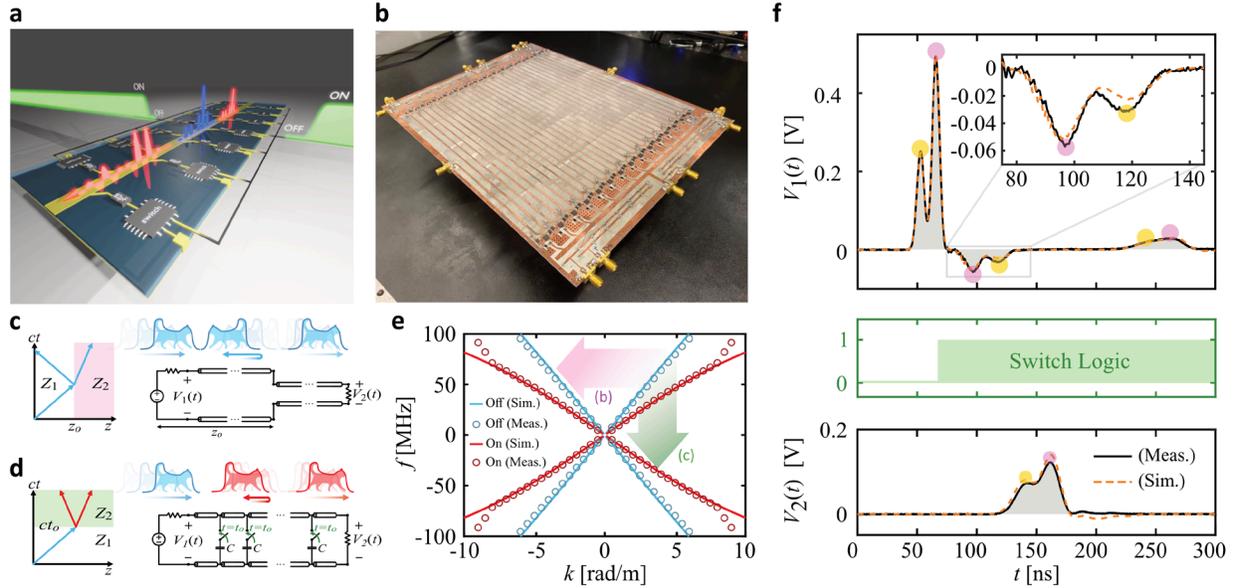

**Fig. 1: Observation of photonic time-reflection. (a)** Illustration of a time-interface in a uniformly switched TLM. A step-like bias signal (in green) is used to uniformly activate a set of switches distributed along the TLM, with spacing much smaller than the wavelengths of operation. Upon closing (opening) of the switches, the effective TLM impedance is abruptly decreased (increased) by a factor of two, causing a broadband forward-propagating signal (in blue) to be split into a time-refracted and a time-reflected signal, both with redshifted frequencies (in red). **(b)** Photo of the fabricated time-switched TLM. **(c)** Reflection at a spatial interface causes the reflected signal to invert its profile in space, while **(d)** a temporal interface breaks time-translation symmetry in a spatially homogeneous medium, uniformly inverting in time the evolution of an input signal. **(e)** Simulated and measured dispersion relation of the fabricated TLM before (blue) and after (red) activating the switches. The purple arrow indicates broadband transitions induced by spatial reflection, coupling positive and negative momenta, while the green arrow indicates TR, coupling positive and negative frequencies. **(f)** Experimental observation of photonic TR at a time-interface with an asymmetric pulse consisting of a smaller input signal (yellow marker in input port voltage $V_1$) followed by a larger one (purple marker). Within 3ns after the switch logic (middle panel) turns on, a portion of the input signal undergoes TR, propagating back to the input port, where the two signals are measured in reverse order, (purple marker, then yellow, see also zoomed inset), with flipped polarity. After $\approx 140$ ns, the time-refracted signal, having undergone a spatial reflection at the end of the TLM, returns to the $V_1$ port in the original yellow-purple order. In the lower panel, the output-port voltage $V_2$ shows the time-refracted signal, broadened in time due to the broadband redshift of its frequency content induced by the time-interface. Signal amplitudes are plotted accounting for the power lost in the splitter and impedance change (see SI Sec. S3).

A photograph of our fabricated TLM is shown in Fig. 1b. A broadband input signal is injected from one of the ports, and it travels along the meandered microstrip line, loaded by an array of 30 switches connected in series to an array of subwavelength-spaced capacitors (unit cell length ≈ 20.1 cm, see SI Secs. S1, S2 for details on the TLM design and implementation). The meandered microstrip emulates an unbounded medium with close-to-linear dispersion and, while the signal is fully within the TLM, a control signal for the switches is sent via a pair of much shorter microstrips (with 80 times faster transit time across the TLM), synchronously triggering all the switches with a rise-time ≈ 3 ns, much faster than the temporal dynamics of the incoming wave (see SI Sec. S8 for details on switch synchronization). This switching event is much faster than half of a wave period, and its amplitude is of order unity, as required for efficient TR[29], resulting in an efficient time-interface.

At a spatial interface (Fig. 1c), translational symmetry is broken, hence the reflected waves undergo parity inversion ($z \rightarrow -z$) and a receiver at the source location registers the features of a reflected signal in the same order as they were originally sent, akin to a sound echo. Frequency is conserved in this scenario, while wavevector and momentum are not. Conversely, at a time-interface (Fig. 1d) time-translation symmetry is broken while spatial symmetry is preserved. Hence, the "echo" associated with the time-reversed ($t \rightarrow -t$) signal is detected backwards, while the signal retains its original spatial profile due to momentum conservation. In addition, the broadband frequency content of the input signal is abruptly transformed, as predicted by the band diagrams of our TLM, shown in Fig. 1e. Blue and red lines depict the TLM dispersion curve before and after the switching, comparing simulated (lines) and measured (circles) results. Given the small spacing between neighboring loads compared to the relevant wavelengths, the curves follow a linear dispersion, with different slopes corresponding to the different effective capacitance before and after the switching. Wave scattering at a spatial discontinuity (Fig. 1c) is equivalent to a horizontal transition in the dispersion diagram (purple arrow in Fig. 1e), preserving frequency and generating waves with new positive and negative momenta. Conversely, a time-interface (Fig. 1d) corresponds to a vertical transition (green arrow in Fig. 1e), which preserves wavenumber and generates new positive and negative frequencies, efficiently translating the entire frequency spectrum of a broadband input wave, while conserving the entire spatial structure of the pulse.

These features are clearly observed in our time-domain experimental measurements (Fig. 1f): we excite the TLM with an input signal consisting of an asymmetric pair of Gaussian pulses, measured by the input port voltage $V_1(t)$ as a first smaller pulse (yellow marker) followed by a larger one (purple marker). Approximately 15 ns after the activation of the switches (middle panel of Fig. 1f, see Methods for details on the timing), we record the time-reflected signal at the input port, whose zoomed-in view reveals it to be the TR-copy (purple → yellow) of the input. This TR-signal has inverted polarity with respect to the input signal, indicating that the TR coefficient is negative, as expected from the scattering coefficients for a reduction in wave impedance achieved by connecting the lumped capacitors in our TLM (see SI Sec. S5 for derivation):

$$R_{1\to 2} = \frac{Z_2(Z_2 - Z_1)}{2Z_1^2}, \qquad (1)$$

where in our system $Z_1 \approx 50\Omega$ is the line impedance before switching and $Z_2 \approx 25\Omega$ is the line impedance after switching. Approximately 140 ns later, an attenuated signal is received at the input port, corresponding to the time-refracted signal that has been traveling to the end of the TLM and then spatially reflected backwards at the mismatched termination. As expected, this second signal has inverted symmetry compared to the TR signal (yellow → purple). The TR signal, as well as the time-refracted signal at the output port $V_2$ (Fig. 1f, bottom panel), retain the same spatial profile as the incident signal due to the preserved spatial symmetry, but slow down, as they travel in a line with increased effective permittivity. This phenomenon underpins the broadband and efficient frequency translation process occurring abruptly at the time-interface, based on which each frequency component of the input signal is transformed according to $\omega_1 \to \omega_2 = (Z_2/Z_1)\omega_1$ (see SI Sec. S5 for derivation).

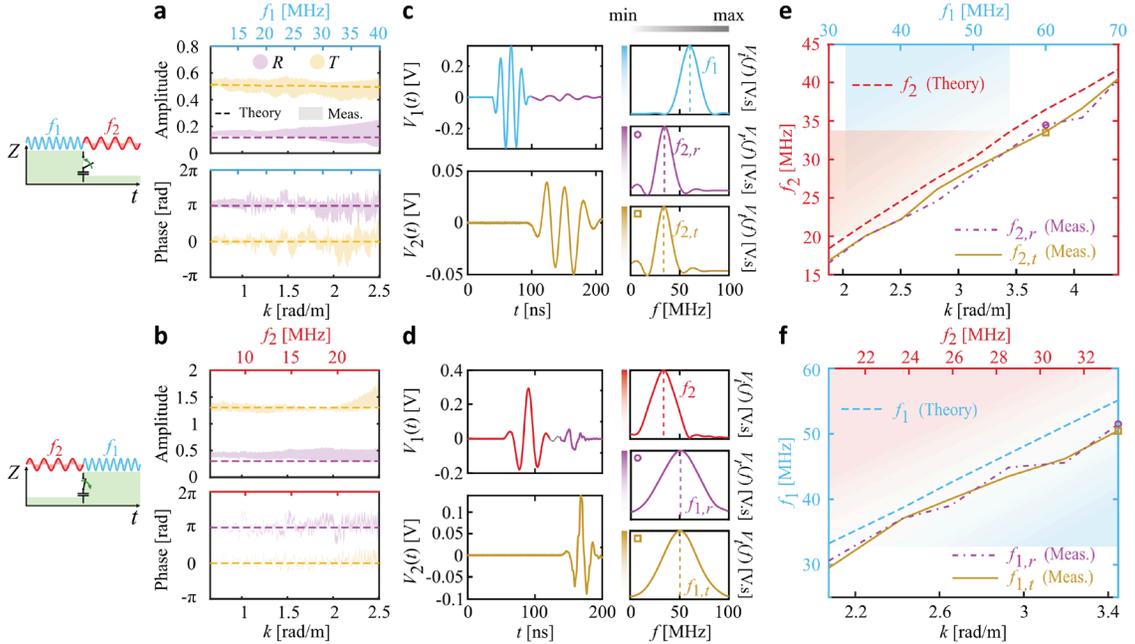

**Fig. 2. Spectral analysis of photonic time-reflection.** Leftmost panels: schematics of negative (top row) and positive (bottom row) switching of the effective impedance. $f_1$ (blue) and $f_2$ (red) denote the wave frequencies when the switches are closed and open respectively. **(a-b)** Amplitude (top) and phase (bottom) of the spectra of the measured (shaded) and theoretical (dotted) time-refraction (T) and TR (R) coefficients upon an impedance (a) decrease and (b) increase, measured from the Fourier transforms of the respective scattered pulses. Note the $\pi$ phase shift upon TR in (b), contrasting with the TR coefficients derived in the literature when switching the medium permittivity. **(c-d)** Carrier-wave measurements of incoming (blue), TR (purple) and time-refracted (yellow) in the time (left column) and frequency (right column) domains, showing broadband (c) redshift and (d) blueshift induced by the time-interface. **(e-**

**f)** Theoretical (dashed) and measured broadband (e) redshift and (f) blueshift of outgoing TR (dot-dashed purple line) and time-refracted (continuous yellow line) signals induced by the time-interface, obtained by scanning the carrier frequency of a narrow-band input signal. Purple circles and yellow squares denote the scenarios in panels (c-d).

In Fig. 2a, we retrieve the TR (R) and time-refraction (T) coefficients by Fourier analyzing the measured output signals $V_1$ and $V_2$ in Fig. 1f, as detailed in the SI Sec. S4. The retrieved amplitudes (top panel) and phases (bottom) of the temporal scattering coefficients as a function of the input wavevector $k$ (lower horizontal axis) and corresponding input frequency (upper horizontal axes), agree well with our theoretical predictions assuming an instantaneous switching event (dashed lines). As expected, the time-refracted signal is in phase with the input, while the TR signal flips sign for all input momenta of our broadband pulse spectrum. The slight wavevector dependence of the scattering coefficients at higher frequencies can be attributed to the finite switching speed (see SI Secs. S9, S10 for an investigation of the switching rise time), the frequency dispersion associated with nonidealities of the circuit components, as well as the finite spacing between neighboring switches.

Our results not only demonstrate efficient TR at a time-interface, but they also imply the evidence of a new form of boundary conditions associated with time-interfaces. Our platform enables fast and efficient impedance changes by adding and removing reactance to the TLM through switches, rather than modifying the reactance in time. Since the involved additional capacitors are static, this operation enables much faster transitions, without heavy requirements on energy, hence addressing the challenges pointed out in Ref. 27. In turn, our time-switched TLM do not necessarily conserve charge at the temporal boundary, different from the common assumption in the existing literature on time-interfaces[1,2,6,8,30]. When closing the switches and connecting the capacitors, we do preserve the total charge in the TLM, which ensures continuity of the displacement field **D**. In contrast, when we open the switches, we abruptly cut-off the charged capacitors from the TLM, creating a charge discontinuity while preserving the voltage. In other words, a new boundary condition needs to be introduced at this type of time interface, in which the electric field **E** and not **D** is conserved. This modified temporal boundary condition leads to new scattering coefficients (see SI Sec. S5):

$$R_{2\to 1} = \frac{Z_2 - Z_1}{2Z_2}, \qquad (2)$$

which is different from Eq. (1). Indeed, our experiments confirm a close agreement between the experimentally retrieved scattering spectra (Fig. 2b) for the charge-discontinuous time-interface and the predictions given by Eq. (2), and unveil the importance of considering the specific dynamics of a temporal interface in order to correctly predict its resulting temporal scattering.

To quantify the broadband nature of frequency translation at our photonic time-interfaces, we carried out temporal scattering experiments with relatively narrow-band input signals (bandwidth $\approx 30\,\text{MHz}$) at time-interfaces featuring increasing (decreasing) impedance, as shown in Fig. 2c (2d). We observe a clear redshift (blueshift) of the carrier frequency from $f_1 = 60$ MHz to $f_{2,r} = 34.5$ MHz and $f_{2,t} = 33.6$ MHz ($f_2 = 33.6$ MHz to $f_{1,r} = 49.5$ MHz and $f_{1,t} = 50.1$ MHz), accompanied by a shrinking (broadening) of the pulse width from $\Delta f_1 = 21$ MHz to $\Delta f_{2,r} = \Delta f_{2,t} = 16.0$ MHz ($\Delta f_2 = 29.1$ MHz to $\Delta f_{1,r} = 48.3$ MHz and $\Delta f_{1,t} = 42.0$ MHz). In Figs. 2e and 2f, we sweep the input carrier wavenumbers $k$ (bottom axes), or equivalently the input frequencies (top axes) for the two switching scenarios, and observe the output frequency (vertical axes). When activating the time-interface by closing the switches and decreasing the wave impedance (Fig. 2e), the center frequency of both TR ($f_{2,r}$) and time-refracted ($f_{2,t}$) waves is redshifted almost uniformly by $\approx 55\%$, over a range of input carrier frequencies spanning the interval 30-70 MHz (see SI Sec. S6 for details on frequency translation measurements). As control experiments, we also measured the frequency up-conversion by the reversed time-interface (Fig. 2f). While scanning the input frequency from 20-34 MHz, the observed blueshift exactly mirrors the process demonstrated by Fig. 2e. We stress that the bandwidth and linearity of the frequency conversion process at a time-interface are only limited by the dispersion of our TLM, as opposed to conventional narrowband frequency conversion processes, opening exciting opportunities in a wide range of photonic applications.

By combining multiple time-interfaces, it is possible to leverage TR-induced interference to manipulate more markedly the input signals. For example, by combining two time-interfaces we form a temporal slab, realized by closing and reopening the switches after a delay. These sequential TR events induce temporal wave interference. Due to causality, scattering phenomena at a temporal slab are markedly different from those in a spatial slab: while multiple scattering occurs in a spatial slab (Fig. 3a), generating a superposition of refracted and reflected waves, a temporal slab produces a total of four scattered waves after the second time-interface (Fig. 3b). In order to experimentally probe the properties of the temporal slab, we launch broadband signals into our TLM (Fig. 3c, black portion of the plots), and observe the total time-scattering (colored portions). For each duration of the temporal slab ($\tau = 15, 25, 35$ ns, see SI Sec S7 for definition of $\tau$), we indeed observe the two expected reflected pulses, each having undergone one TR and one time-refraction process, in opposite order. Notably, as opposed to conventional photonic time-interfaces[1,2,8] and recent theoretical work on temporal slabs[6,30], both TR-signals here are out of phase with respect to the input signal, due to the different temporal boundary condition discussed above, and as shown in the phase spectra in Fig. 2a and 2b. Importantly, all scattered pulses have approximately the same duration as the input signal, since the second time-interface converts back the frequency spectrum to the original frequencies, corresponding to two opposite vertical transitions in the dispersion diagram.

As it can be observed in Fig. 3c, the time delay between the consecutive scattered pulses is proportional to the slab duration. This suggests that the temporal slab can be tuned to control wave interference, thereby realizing the temporal analogue of a Fabry-Perot etalon which enables accurate shaping of the output frequency content. In order to demonstrate this effect explicitly (see also SI Sec. S7), we take the Fourier transform of the TR signal $V_r(t)$, and normalize it against the transform of the input signal $V_i(t)$. This reveals the total reflection spectra (Fig. 3d) of the different temporal slabs as functions of the input wavenumber $k$ (bottom horizontal axis), or equivalently, the input frequency $f_1$ (top axis). For each slab duration, specific values of $k$ feature zero reflection, due to destructive temporal interference between two TR waves, in analogy with the reflection zeros of a Fabry-Perot cavity. In this case, however, the associated phase accumulation does not occur in space, but rather in time. To further highlight the versatile spectral control offered by our temporal slab through temporal interference, in Fig. 3e we examine the total TR at a fixed input wavenumber $k = 1.8$ rad/m (dashed vertical line in Fig. 3d), as we increase the slab duration, for six pulses of different half-maximum durations ranging from 5 ns to 10 ns, comparing measured and theoretical results. We observe how the total amplitude of the time-reversed waves can be continuously tuned by varying $\tau$, granting us dynamic control over wave interference at the two time-interfaces without having to change the lumped capacitance values. In SI Sec. S11, we also consider the case of an inverse slab, whereby the switches are first opened and then closed again.

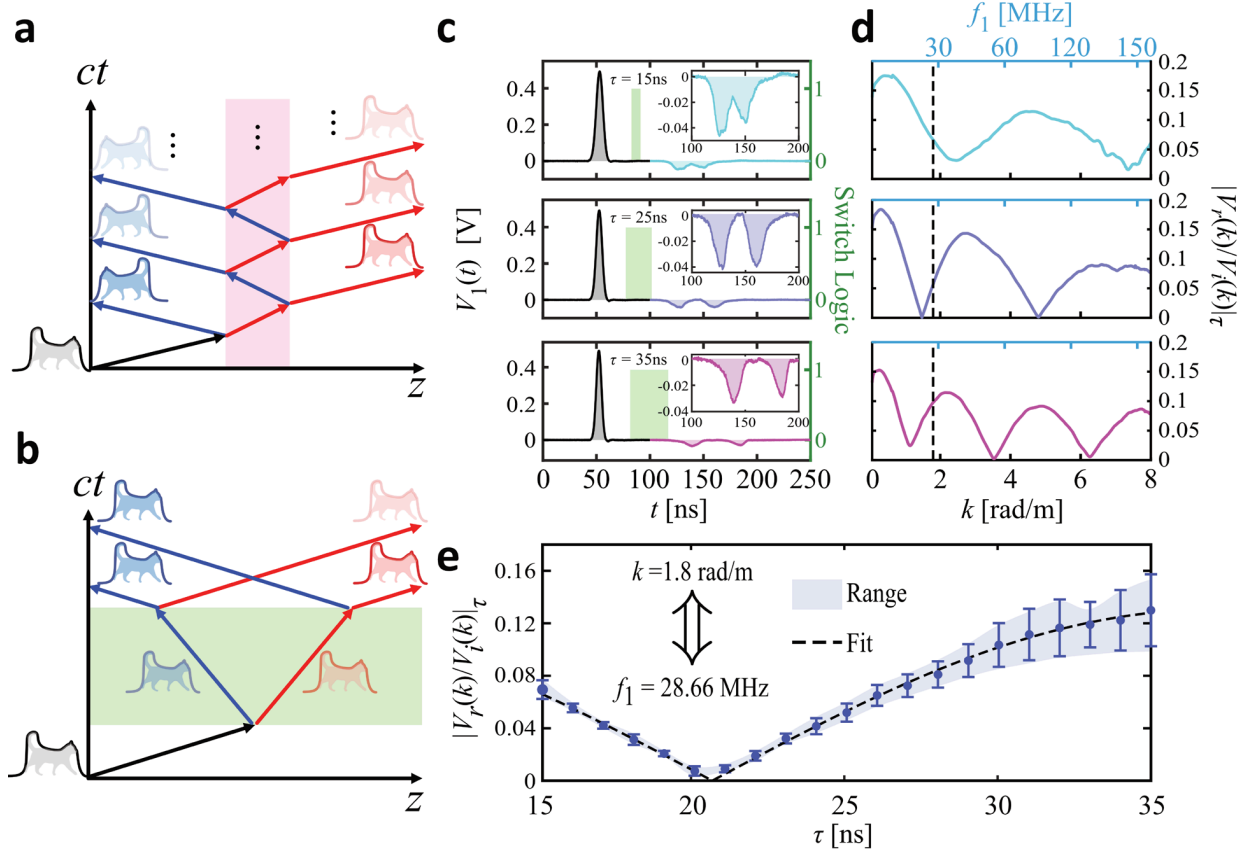

**Fig. 3. Wave scattering from a temporal slab.** **(a)** Conceptual sketches of a spatial and **(b)** a temporal slab with stepped wave impedance. Inside a spatial slab, multiple partial reflections occur, gradually decaying with increasing scattering orders, whereas at a temporal slab only four scattered waves interfere with each other. **(c)** Experimentally measured voltages at input ($V_1$) and output ($V_2$) ports after temporal slabs with varying durations. Here, $\tau \in [15 \text{ ns}, 25 \text{ ns}, 35 \text{ ns}]$ is the "ON" time of the control signal, and the corresponding logic states for the switches are indicated by the green blocks. In each plot, the double reflection induced by the slab (cyan, purple and magenta portions) are clearly visible. The elapsed time between reflected pulses is proportional to the slab duration. **(d)** Normalized amplitudes of the total TR signals as a function of wavenumber $k$ for the different temporal slabs of panel c, exhibiting zero reflection at selected frequencies, controlled by the slab duration. **(e)** For fixed $k$ (dot-dashed vertical line in panel d), we measure the normalized reflection amplitude as a function of switching duration $\tau$, demonstrating large, continuous spectral tunability.

To conclude, in this work we have reported the first observation of photonic time-reflection for broadband and efficient phase conjugation and frequency translation at single and double time-interfaces, as well as the demonstration of controlled time-reversal-induced interference phenomena from a temporal slab formed by a pair of time-interfaces. These results establish the key building blocks towards the realization of time-metamaterials and photonic time crystals, opening a wide range of opportunities in the rising field of time-varying photonic media[7,8,28], with applications for ultrafast wave routing[24] and focusing[3,21], negative refraction[31,32], efficient and broadband spectral filtering and frequency manipulation[33], novel forms of ultrafast energy mixing[34,35], and photonic Floquet matter[4,12]. Our approach to realize time-interfaces using external reactive elements added and removed through switches is key to these demonstrations, and it can be straightforwardly configured to introduce at the same time spatial and temporal interfaces, for instance by activating only a portion of the switches, thus blending together spatial and temporal degrees of freedom and enabling even more flexibility in wave control and manipulation. Field-programmable gate arrays controlling the switches may realize real-time reconfigurability and even self-adaptation of the response.

More broadly, our results open a pathway to employ time-interfaces for broadband, efficient phase conjugation and frequency conversion arising over very short time scales, of great relevance for a variety of applications in electromagnetics and photonics. Efficient time-reversal is important in the context of wireless communications and radar technologies, for instance for channel estimation, which is now performed through complex digital computations, to estimate complex propagation channels and compensate for signal distortion and dispersion. Broadband efficient frequency conversion is also key for applications spanning from night-vision systems to quantum photonics. Of particular interest would be to extend these concepts to higher frequencies, and we envision several available routes, from modern CMOS technology, which may deliver switching speeds up to 2 orders of magnitude faster than those reported here, extending the frequency range to the THz regime, as well as all-optical approaches leveraging giant nonlinearities in graphene, which offer low switching power and times as short as 260 fs[36], or flash ionization in plasmas with even shorter switching times[37].

**Acknowledgments**

This work was supported by the Air Force Office of Scientific Research MURI program with grant No. FA9550-18-1-0379, the Simons Collaboration on Extreme Wave Phenomena and a Vannevar Bush Faculty Fellowship. E.G. was supported by the Simons Foundation through a Junior Fellowship of the Simons Society of Fellows (855344,EG).

# References

1. Morgenthaler, R. Velocity Modulation of Electromagnetic Waves. *IRE Trans. Microw. Theory Tech.* 167–172 (1958). doi:10.1109/TMTT.1958.1124533

2. Fante, R. L. Transmission of Electromagnetic Waves into Time-Varying Media. *IEEE Trans. Antennas Propag.* **AP-19,** 417–424 (1971).

3. Bacot, V., Labousse, M., Eddi, A., Fink, M. & Fort, E. Time reversal and holography with spacetime transformations. *Nat. Phys.* **12,** 972–977 (2016).

4. Yin, S., Galiffi, E. & Alù, A. Floquet metamaterials. *eLight* **2,** 8 (2022).

5. Engheta, N. Metamaterials with high degrees of freedom: Space, time, and more. *Nanophotonics* **10,** 639–642 (2021).

6. Pacheco-Peña, V. & Engheta, N. Antireflection temporal coatings. *Optica* **7,** 323–331 (2020).

7. Caloz, C. & Deck-Leger, Z.-L. Spacetime Metamaterials—Part I: General Concepts. *IEEE Trans. Antennas Propag.* **68,** 1569–1582 (2020).

8. Caloz, C. & Deck-Leger, Z.-L. Spacetime Metamaterials—Part II: Theory and Applications. *IEEE Trans. Antennas Propag.* **68,** 1583–1598 (2020).

9. Lustig, E., Segev, M. & Sharabi, Y. Topological aspects of photonic time crystals. *Opt. Vol 5 Issue 11 Pp 1390-1395* **5,** 1390–1395 (2018).

10. Lyubarov, M., Lumer, Y., Dikopoltsev, A., Lustig, E., Sharabi, Y. & Segev, M. Amplified emission and lasing in photonic time crystals. *Science* **0,** eabo3324 (2022).

11. Winn, J. N., Fan, S., Joannopoulos, J. D. & Ippen, E. P. Interband transitions in photonic crystals. *Phys. Rev. B* **59,** 1551–1554 (1999).

12. Dutt, A., Lin, Q., Yuan, L., Minkov, M., Xiao, M. & Fan, S. A single photonic cavity with two independent physical synthetic dimensions. *Science* **367,** 59–64 (2020).

13. Fleury, R., Khanikaev, A. B. & Alù, A. Floquet topological insulators for sound. *Nat. Commun.* **7,** 1–11 (2016).

14. Cartella, A., Nova, T. F., Fechner, M., Merlin, R. & Cavalleri, A. Parametric amplification of optical phonons. *Proc. Natl. Acad. Sci. U. S. A.* **115,** 12148–12151 (2018).

15. Shan, J.-Y., Ye, M., Chu, H., Lee, S., Park, J.-G., Balents, L. & Hsieh, D. Giant modulation of optical nonlinearity by Floquet engineering. *Nature* **600,** 235–239 (2021).

16. Hayran, Z., Chen, A. & Monticone, F. Spectral causality and the scattering of waves. *Optica* **8,** 1040 (2021).

17. Lerosey, G., de Rosny, J., Tourin, A., Derode, A., Montaldo, G. & Fink, M. Time Reversal of Electromagnetic Waves. *Phys. Rev. Lett.* **92,** 193904 (2004).

18. Fink, M. & Prada, C. Acoustic time-reversal mirrors. *Inverse Probl.* **17,** R1–R38 (2001).

19. Chumak, A. V., Tiberkevich, V. S., Karenowska, A. D., Serga, A. A., Gregg, J. F., Slavin, A. N. & Hillebrands, B. All-linear time reversal by a dynamic artificial crystal. *Nat. Commun.* **1,** 141 (2010).

20. Vezzoli, S., Bruno, V., Devault, C., Roger, T., Shalaev, V. M., Boltasseva, A., Ferrera, M., Clerici, M., Dubietis, A. & Faccio, D. Optical Time Reversal from Time-Dependent Epsilon-Near-Zero Media. *Phys. Rev. Lett.* **120,** 43902 (2018).

21. Lerosey, G., De Rosny, J., Tourin, A., & Fink, M. Focusing beyond the diffraction limit with far-field time reversal. *Science*, *315*(5815), 1120-1122 (2007).

22. Mosk, A. P., Lagendijk, A., Lerosey, G., & Fink, M. (2012). Controlling waves in space and time for imaging and focusing in complex media. *Nature photonics*, *6*(5), 283-292.

23. Akbarzadeh, A., Chamanara, N. & Caloz, C. Inverse prism based on temporal discontinuity and spatial dispersion. *Opt. Lett.* **43,** (2018).

24. Pacheco-Peña, V. & Engheta, N. Temporal aiming. *Light Sci. Appl.* 9, 1–12 (2020).

25. Rizza, C., Castaldi, G., and Galdi, V. Short-Pulsed Metamaterials *Phys. Rev. Lett*. **128**, 257402 (2022).

26. Liberal, I., & Vázquez-Lozano, J. E. Shaping the quantum vacuum with anisotropic temporal boundaries. *arXiv preprint arXiv:2208.10776* (2022).

27. Hayran, Z., Khurgin, J. B., & Monticone, F. ℏω versus ℏk: dispersion and energy constraints on time-varying photonic materials and time crystals. *Opt. Mat. Expr.* **12**, 3904 (2022).

28. Galiffi, E., Tirole, R., Yin, S., Li, H., Vezzoli, S., Huidobro, P. A., Silveirinha, M. G., Sapienza, R., Alù, A. & Pendry, J. B. Photonics of time-varying media. *Adv. Photonics* **4,** 014002 (2022).

29. Galiffi, E., Yin, S. & Alú, A. Tapered photonic switching. *Nanophotonics* **11**, 3575–3581 (2022).

30. Mendonça, J. T., Martins, A. M. & Guerreiro, A. Temporal beam splitter and temporal interference. *Phys. Rev. A* **68,** 043801 (2003).

31. Pendry, J. B. Time Reversal and Negative Refraction. *Science* **322,** 71–73 (2008).

32. Bruno, V., Devault, C., Vezzoli, S., Kudyshev, Z., Huq, T., Mignuzzi, S., Jacassi, A., Saha, S., Shah, Y. D., Maier, S. A., Cumming, D. R. S., Boltasseva, A., Ferrera, M., Clerici, M., Faccio, D., Sapienza, R. & Shalaev, V. M. Negative Refraction in Time-Varying Strongly Coupled Plasmonic-Antenna- Epsilon-Near-Zero Systems. *Phys. Rev. Lett.* **124,** 43902 (2020).

33. Miyamaru, F., Mizuo, C., Nakanishi, T., Nakata, Y., Hasebe, K., Nagase, S., Matsubara, Y., Goto, Y., Pérez-Urquizo, J., Madéo, J., Dani, K. M., Joel, P.-U., Madéo, J. & Dani, K. M.
11

Ultrafast Frequency-Shift Dynamics at Temporal Boundary Induced by Structural-Dispersion Switching of Waveguides. *Phys. Rev. Lett.* **127,** 053902 (2021).

34. Li, H., Yin, S., Galiffi, E. & Alù, A. Temporal Parity-Time Symmetry for Extreme Energy Transformations. *Phys. Rev. Lett.* **127,** 153903 (2021).

35. Carminati, R., Chen, H., Pierrat, R. & Shapiro, B. Universal statistics of waves in a random time-varying medium. *Phys. Rev. Lett.* **127,** 94101 (2021).

36. Ono, M., Hata, M., Tsunekawa, M., Nozaki, K., Sumikura, H., Chiba, H., & Notomi, M. (2020). Ultrafast and energy-efficient all-optical switching with graphene-loaded deep-subwavelength plasmonic waveguides. Nature Photonics, 14(1), 37-43.

37. Nishida, A., Yugami, N., Higashiguchi, T., Otsuka, T., Suzuki, F., Nakata, M., Sentoku, Y. & Kodama, R. (2012). Experimental observation of frequency up-conversion by flash ionization. Applied Physics Letters, 101(16), 161118.

# Methods
## Simulations
All circuit simulations are performed in Keysight Advanced Design Systems[S1]. Time domain simulations are carried out using the Transient solver, while frequency domain analysis is done with the S-Parameters solver.

The main TLM carrying the signal is modelled with physical transmission line sections with characteristic impedance $Z_o = 50\ \Omega$, length $d = 0.208$ m, effective dielectric constant $\epsilon_{r,eff} = 8.36$, attenuation constant $\alpha = 0.5$ (at $f = 100$ MHz, dielectric loss tangent $\tan \delta = 0.0019$. It should be noted that the effective dielectric constant of 8.36 is consistent with expectations from well-known design formulas with microstrip lines. To further model the non-idealities associated with our manufactured circuit, each unit cell is connected to its nearest with a 2 pF capacitor, and to its second nearest neighbor with a 1 pF capacitor. This models the parasitic capacitive coupling between the unit cells, which limits our operating bandwidth by introducing dispersion and bandgaps. These values are found to produce consistently good match between experimental and simulation results.

The switches are implemented as ideal voltage controlled reflective switches with $1\ \Omega$ resistance in the "on" state. Each switch connects a node on the transmission line to a series RLC circuit, with an 82 pF capacitor representing the load, a $4\ \Omega$ resistor representing the parasitic losses, and an 8 nH inductor representing the parasitic inductance of the grounding vias. These values are extracted from measurements of a single unit cell: the parasitic component values are swept in frequency domain simulations until a good match with the measured S-parameters is observed.

## Experiments
The TLM sample is fabricated in house using a 1.52 mm thick Rogers TMM 13i substrate, which has a nominal dielectric constant of $12.85 \pm 0.35$, and a loss tangent of 0.0019. See SI Secs. S1, S2 for details of the circuit layout and choices of all circuit components.

During our measurements, the input signal consists of a train of repeated copies of an arbitrarily shaped pulse. Since the input pulses generally have durations in the order of tens of ns, which is much shorter than the period of the pulse train ($> 1\mu s$), we can effectively treat each pulse as a single isolated event. The signals entering the TLM through the input port is probed with a T-connector attached to an oscilloscope, while the exiting signals are probed directly with the scope. See SI Sec. S2 for a simplified schematic of the experimental setup.

To control the switches, we use a rectangular pulse train that is phased locked to the input pulse train. By adjusting the relative phase between the two signals, we can control the timing of the interface. Notably, the switches that we used have internal grounding circuitry which, upon opening, will dissipate all charges stored on the capacitor[S3]. This is crucial for the accurate realizing of our two distinct types of time-interfaces.

To induce the temporal reflection, we activate the switches after the input signal is registered in $V_{o1}(t)$, but before it is recorded in $V_{o2}(t)$, corresponding to a time when the signal is completely contained within the TLM. The time-reflected signal will travel backwards towards Port 1 of the transmission line, where it will again enter the input T-connector. Identical copies of the time-reflection (with 3.5 dB attenuation) are fed to the source and the oscilloscope. Hence, the actual amplitude of the temporal reflection generated by our time-interface is larger than that recorded by Port 1. See Supplementary Information for the effect of the T-connector and impedance mismatch.

To measure the frequency translation of our time-interface, we launched signals with varying carrier frequencies into the transmission-line, and observed the time-reflected as well as refracted signals. See SI Secs. S6, S12 for a typical time-domain measurement result, and the accounting of video leakage. We performed Fourier analysis on the time-gated incident, refracted, and reflected signals to observe their individual frequency content. The frequency spectra for each signal will consist of a sharp peak, aligned with its center frequency.

# Supplementary Information

## S1. Theoretical Design and Optimization of the Loaded Transmission Line Metamaterial

Each unit cell of our transmission line, once the switches have been activated, can be approximately modelled by the T-network shown in Fig. S1, where $d$ is the physical length of the cell (i.e. one turn of the meandered microstrip on our printed circuit board), and $jb = j\omega C Z_o$ is the susceptance of the capacitive load, normalized against $Z_o$, the characteristic impedance of the unloaded TLM. The voltage and currents at the input port of such a network can be related to those at the output port by its ABCD matrix, which reads[S2],

$$\begin{bmatrix} A & B \\ C & D \end{bmatrix} = \begin{bmatrix} \left(\cos\theta - \frac{b}{2}\sin\theta\right) & j\left(\sin\theta + \frac{b}{2}\cos\theta - \frac{b}{2}\right) \\ j\left(\sin\theta + \frac{b}{2}\cos\theta + \frac{b}{2}\right) & \left(\cos\theta - \frac{b}{2}\sin\theta\right) \end{bmatrix}. \tag{S1}$$

Here, $\theta = k_o d$ is the total propagation phase through the unloaded transmission-line section. Applying Bloch boundary condition yields the dispersion relation for an infinitely long chain of this unit cell as

$$\cos(\beta d) = \cos\theta - \frac{b}{2}\sin\theta. \tag{S2}$$

Furthermore, the Bloch wave impedance of the periodic structure is given by

$$Z_B = \pm \frac{B Z_o}{\sqrt{A^2 - 1}}. \tag{S3}$$

In Fig. S2, we plot the normalized dispersion relation and the Bloch wave impedance of our TLM, for $\theta = 72°$ and $b = 2.753$. where $\omega_o = 2\pi \times 100$ MHz is the designated maximum operating frequency. In this case, the loaded TLM to exhibit a Bloch impedance is exactly $Z_o/2$ at $\omega = \omega_o/2 = 2\pi \times 50$ MHz, and retains approximately the same value for lower frequencies.

Having determined the ideal frequency translation ratio $f_2/f_1$, as well as the wave impedances before ($Z_1$) and after ($Z_2$) switching, one can calculate the theoretical temporal reflection and refraction coefficient of the TLM. For generality, we present a set of parametric studies that were conducted to identify the best design parameters for the TLM, which can be scaled to accommodate different specifications (i.e., different temporal reflection magnitude and frequency translation ratios).

At the design stage, two key parameters need to be determined: the length of the unit cell, which is directly proportional to the phase accumulation; and the appropriate shunt load $b$. Both parameters will have significant impact on various aspects of the time-interface. Crucially, together, they determine the band gaps of our TLM. Since we wish to operate in a regime in which the TLM is almost translationally homogeneous and dispersionless, we must pay special attention to avoid the first band gap. To this end, we perform parametric studies on $R$ and $f_2/f_1$ by sweeping $\theta$ and $b$ (which is equivalent to sweeping the cutoff frequency of the first band gap, $\omega_c$); the results for $R$ and $f_2/f_1$ are plotted in Fig. S3A and Fig. S3B respectively. In Fig. S3A, we use a red line to denote the combinations of $(\theta, \omega_c)$ that gives the maximum temporal reflection magnitude.

## S2. Practical Implementation

The practical circuit layout is shown in Fig. S4 and the choices of all circuit components are listed in Table **S1**. A schematic of our experimental set up for time domain measurements is shown in Fig. S5. The input signal is launched from an arbitrary waveform generator (Keysight M8195A) into a symmetric T-connector. Identical copies of the input (both with 3.5 dB attenuation) are fed to Port 1 of the transmission line as well as channel 1 of an oscilloscope (Tektronix DPO4104), which records $V_1(t)$. Port 2 of the transmission line is connected to channel 2 of the oscilloscope, which records $V_2(t)$. Both oscilloscope channels have an input impedance of 50 Ω, and record in the averaged mode with 512 samples.

## S3. Compensation of the T-connectors and Impedance Mismatch

We measured the signal via an oscilloscope at the Port 1 at $x = 0$, denoted as $V_{o1}(t)$, which is a linear superposition of input and time-reflected signals. Considering the effect of the T-connector that connects the source (denoted by subscript "$s$"), transmission-line (subscript "$t$"), and the oscilloscope (subscript "$o$"), we can estimate $V_{o1}(t)$ for the case of closing switches as

$$V_{o1}(t) \approx \mathcal{F}^{-1}\left\{S_{os}(\omega)\frac{\mathcal{F}[V_{inc}(0,t)]}{S_{ts}(\omega)} + S_{ot}(\omega)(1+\Gamma)\mathcal{F}[V_{ref}(0,t)]\right\}, \quad (S4)$$

where $\mathcal{F}$ denotes the Fourier transform. Then $S$-parameters measurement of the T-connector gives $S_{os} \approx S_{ts} \approx S_{ot} \approx 0.665$ from 10~100 MHz. $\Gamma$ is the spatial reflection coefficient considering the impedance mismatch between the switched TLM and the T-connector, which reads,

$$\Gamma = \frac{Z_0 - Z_{TL}}{Z_0 + Z_{TL}}, \quad (S5)$$

where $Z_{TL} = Z_1 (\approx 50\ \Omega = Z_0)$ with all switches open, while $Z_{TL} = Z_2$ if all switches are closed. The effective impedance of the TLM can be retrieved from its measured $S$-parameters. Similar analysis can be done for other quantities, e.g., that are needed in the main text.

## S4. Retrieval of the Time-Reflection and Time-Refraction Coefficients

The reflection coefficient at a time-interface at $t = t_s$ is defined as

$$R(k) = \frac{\tilde{V}_{ref}(k, t_s^+)}{\tilde{V}_{inc}(k, t_s^-)} = \frac{\int_{-\infty}^{+\infty} V_{ref}(x, t_s^+)e^{jkx}dx}{\int_{-\infty}^{+\infty} V_{inc}(x, t_s^-)e^{jkx}dx}, \quad (S6)$$

where $k$ is the real part of the wavenumber conserved across the time-interface. $\tilde{V}_{inc,ref}(k,t)$ refer to the incident and reflected signals in the momentum space, while $V_{inc,ref}(x,t)$ are the voltages expressed in the real space. We assume that the dispersion and the loss of the transmission line are small enough: $\omega \approx vk$ and then $V_{inc}(x,t) \approx V_{inc}\left(0, t - \frac{x}{v_1}\right)e^{-\alpha_1 x}$, $V_{ref}(x,t) \approx V_{ref}\left(0, t + \frac{x}{v_2}\right)e^{\alpha_2 x}$. The loss rates $\alpha_1$ and $\alpha_2$ can be retrieved by measuring the S-parameters of the unloaded and loaded transmission lines, respectively. By taking Fourier transform of the incident and reflected signals recorded at the input port, one can calculate the reflection coefficient measured at the input port for a given wavenumber $k$ and corresponding frequencies $\omega_{1,2}$:

$$\tilde{R}(k) = \frac{\mathcal{F}[V_{ref}(0,t)]|_{\omega_2}}{\mathcal{F}[V_{inc}(0,t)]|_{\omega_1}} \approx \frac{e^{-jkx}\int_{-\infty}^{+\infty} V_{ref}\left(0, t + \frac{x}{v_2}\right)e^{-j\omega_2 t}dt}{e^{jkx}\int_{-\infty}^{+\infty} V_{inc}\left(0, t - \frac{x}{v_1}\right)e^{-j\omega_1 t}dt} \approx e^{-j2kx}e^{-(\alpha_1+\alpha_2)x}\frac{\int_{-\infty}^{+\infty} V_{ref}(x,t)e^{-j\omega_2 t}dt}{\int_{-\infty}^{+\infty} V_{inc}(x,t)e^{-j\omega_1 t}dt}. \quad (S7)$$

To obtain the exact temporal reflection coefficient, we need to compensate both the phase and loss during the round trip that a wave travels. Without loss of generality, we assume $t_s = 0$. Then, by changing the integration variable $t = -x_r/v_1$ for the denominator and $t = -x_r/v_2$ for the numerator in Eq. (S7), we obtain the relation between the corrected reflection coefficient and that measured at the input port:

$$R(k) \approx \frac{\omega_2}{\omega_1}\tilde{R}(k)e^{j2kx_r}e^{(\alpha_1+\alpha_2)x_r}, \quad (S8)$$

Here, $x_r$ entails the physical distance that a wave travels from the input port until it gets reflected by the time-interface, which can be retrieved via $x_r \approx -\frac{1}{2}\partial_k[\arg(\tilde{R})]$.

Therefore, the procedure for retrieving the time-reflection coefficients for a given wavenumber $k$ can be summarized as follows:
a) Measure the S-parameters and the dispersion diagrams of unloaded and loaded transmission lines.
b) Read the corresponding frequencies $\omega_{1,2}$; Retrieve the loss rate $\alpha_{1,2}$ from the S-parameters.
c) Calculate the reflection coefficient at the input port $\tilde{R}(k)$ using Eq. (S7).
d) Compensate the loss and accumulated phase using Eq. (S8).
Similar procedure also applies to retrieving the time-refraction coefficient, where the relation between the time-refraction coefficient $\tilde{T}(k)$ referring to the two ports and the actual temporal refraction coefficient $T(k)$ reads:

$$T(k) \approx \frac{\omega_2}{\omega_1}\tilde{T}(k,x)e^{jkL}e^{(\alpha_1 x_r + \alpha_2 x_t)}, \quad (S9)$$

where $L = x_r + x_t$ is approximately the total length of our transmission-line, and $x_t$ denotes the retrieved distance that the wave travels after the time-interface until it exits the transmission-line.

## S5. Derivation of the Temporal Scattering Coefficients

Consider a homogeneous transmission-line (TL) whose distributed capacitance can be changed instantaneously at $t = 0$, by closing or opening a switch connected to a shunt capacitor $C_p$, as shown in Fig. S7. The values of the capacitance are chosen such that the effective distributed capacitance of the TL stays at $C_1$ for $t < 0$ while it jumps to $C_2$ for $t > 0$. Applying a Laplace transform $\mathcal{L}\{v(t)\} = \int_0^{+\infty} v(t)e^{-st}dt = V(s)$ on both voltage $v(z,t)$ and current $i(v,t)$, one can write the wave equation for $t > 0$ as:

$$(\partial_z^2 - s^2 L_1 C_2)V = -L_1 C_1 (s + j\omega_1)v(z, 0^-). \tag{S10}$$

Assuming a harmonic variation $e^{-jkz}$ of the fields in space, we can solve for the voltage in $s$-domain. For connecting a parallel capacitor as shown in Fig. S7A,

$$V(z,s) = \frac{C_1}{C_2} \frac{s + j\omega_1}{s^2 + \omega_2^2} v(z, 0^-), \tag{S11}$$

while for the case of opening the switch as shown in Fig. S7B,

$$V(z,s) = \frac{s + j\frac{C_1}{C_2}\omega_1}{s^2 + \omega_2^2} v(z, 0^-). \tag{S12}$$

In both cases, $\omega_2$ is not only the eigenfrequency of the TL after switching, but also relates to the poles of the fields in $s$-domain. It reads:

$$\omega_2 = \omega_1 \sqrt{C_1/C_2} = \omega_1 Z_2/Z_1, \tag{S13}$$

which predicts the theoretical frequency translation ratio.

More interestingly, by applying the initial-value theorem, Eqs. (S11) and (S12) will entail different temporal boundary conditions: when closing the switch (Fig. S7A) the voltage is discontinuous and the charge stored in capacitors is conserved:

$$C_2 v(z, 0^+) = C_1 v(z, 0^-), \tag{S14}$$

while in the case where the switch is opened (Fig. S7B), the voltage remains continuous, yet the charges are not because $C_p$ is suddenly disconnected from the TL system:

$$v(z, 0^+) = v(z, 0^-). \tag{S15}$$

Numerical examples of these two cases are demonstrated in Fig. S7C and D: when connecting $C_p$ such that $C_2 = 4C_1$, the voltage on the TL is discontinuous at $t = 0$, while the total charge in the system is conserved, though part of the charge stored in $C_1$ is transferred to $C_p$ instantaneously.

The frequency is halved as expected. By contrast, when disconnecting $C_p$ from the TL such that $C_2 = C_1/4$. The voltage is continuous across the time-interface, while the total charge decreases as we exclude $C_p$ from the system. The frequency doubles as expected in this case.

Having derived these boundary conditions, we can readily find the temporal reflection and transmission coefficients for switching on a parallel capacitor:

$$T_{on} = \frac{1}{2}\left(\frac{C_1}{C_2} + \sqrt{\frac{C_1}{C_2}}\right) = \frac{Z_{on}(Z_{on} + Z_{off})}{2Z_{off}^2},$$

$$R_{on} = \frac{1}{2}\left(\frac{C_1}{C_2} - \sqrt{\frac{C_1}{C_2}}\right) = \frac{Z_{on}(Z_{on} - Z_{off})}{2Z_{off}^2};$$

(S16)

and for switching off a parallel capacitor:

$$T_{off} = \frac{1}{2}\left(1 + \sqrt{\frac{C_1}{C_2}}\right) = \frac{Z_{on} + Z_{off}}{2Z_{on}},$$

$$R_{off} = \frac{1}{2}\left(1 - \sqrt{\frac{C_1}{C_2}}\right) = \frac{Z_{on} - Z_{off}}{2Z_{on}}.$$

(S17)

In our experiments with temporal slabs, we observed two time-reflected signals, and both are out of phase with respect to the incident signal. This can only be explained via Eqs. (S16) and (S17): in fact, in our setup the effective impedance is decreased when the loaded capacitors are switched on, i.e., $Z_{on} < Z_{off}$, and vice versa. Both temporal reflection coefficients $R_{on,off}$ will be negative, hence, both time-reflected waves ($R_{on}T_{off}$ and $T_{on}R_{off}$) will be out of phase to the incidence, while the two time-refracted waves ($T_{on}T_{off}$ and $R_{on}R_{off}$) will be in phase.

## S6. Measurement of Frequency Conversion with Wave Packets

While determining the broadband frequency translation properties of our TLM, we inject wave packets with different center frequencies, and observe the converted center frequency of the output. Fig. S6 shows a sample of the time-domain measurement used to quantify the frequency conversion of our time-interface when the wave impedance is abruptly reduced via the closing of the switches. The top left plot of Fig. S6 shows the incident (black) and the reflected (red) signals. Note that the effects of the T-connectors and impedance mismatch have not been compensated as they are not crucial for this study. The bottom plot shows the transmitted (blue) signal. Visually, it is clear that the frequencies of the time-scattered waves are lower than that of the incident wave, as the period appears to be longer. We perform Fourier transform on all three signal components. As revealed by the spectra on the right, the original signal has a center frequency of 60 MHz, while the time-reflected and the time-transmitted signals have center frequencies of approximately 35 MHz, corresponding to a down conversion ratio of about 0.55.

## S7. Temporal Slab Acting on a Wave Packet

An ideal temporal slab, formed with two sequential time-interfaces must be unbounded in space. We cannot emulate such a system effectively with our fabricated TLM unless the injected wave packets have very narrow durations (i.e. wide bandwidth). Otherwise, the time-reflections and refractions cannot be completely contained inside the TLM as the temporal slab is being actuated.

In order to study the effect of a temporal slab on a true narrow-band wave packet, we numerically simulate a circuit constructed with three cascaded copies of our TLM. For some arbitrary value of "ON" duration ($\tau = 15$ns), we sweep the center frequency ($f_1$) of the injected pulse (17.5MHz spectral FWHM), and observe the total time reflection (which consists of two interfering wave packets) in both the time domain and the frequency domain. The results for $f_1 \in [30, 35, 38, 41, 46]$ MHz are plotted in Fig. S8. In the left column, we plot the input voltage wave form. The total time reflection induced by the temporal slab for each case is plotted in the middle column. Here, the interference between the two TR pulses are clearly visible. A Fourier transform of the reflected output yields the frequency spectra in the right column. Here, a reflection null persists at approximately 38MHz. The frequency of this null, inversely proportional to the slab duration[S4], is consistent with the experimentally measured value in Fig. 3d of the main text.

The study done in this section also hints at the scalability of our printed circuit board TLM platform: connecting multiple modular samples opens up the opportunity to implement more complex forms of time metamaterials.

## S8. Spatial Homogeneity of the Time-Interface

In our experiment, the control signal for the switches were carried by a pair of transmission lines with very short but nonetheless finite electrical length. Hence, strictly speaking, the switches are not perfectly synchronized. This may introduce some degree of spatial inhomogeneity and serve to detune the wave momentum. However, the control TL is much shorter than the main TL (by more than 40 times). This means that in the perspectives of a pulse propagating on the main TL, the change in the distributed capacitance appears effectively homogeneous. To further enhance the spatial homogeneity, we feed the control signal from the center of the sample, instead of the end. This serves to reduce the total transit time of the control signal across the entire sample by another factor of two, compared to the transit time of the signal on the main TL.

To prove the above assertion, we perform circuit simulations with different degrees of switch desynchronization; the results are shown in Fig. S9. Here, we inject an asymmetric signal in the same form as that used in Fig. 2e of the main text. The grey curves depict the voltages at port 1 and 2 of the TLM if we turn on all of the switches in a perfectly synchronized fashion. In other words, there is no time delay between the activation of adjacent switches. The dashed red curve corresponds to the simulated results when the electrical lengths between adjacent switches ($L_c$) match our experiment exactly (<1cm). It can be seen that the red curves overlap the ideal grey curves almost exactly, showing that our experimental platform can realize an effectively homogeneous time-interface that can almost conserve momentum for sufficiently low frequencies.

## S9. Characterization of the Circuit Switching Speed

As pointed out by the main text as well as previous works in the literature[S5,S6], in order to induce strong temporal reflections, the time-interface must be sufficiently fast. Ideally, the duration of the interface should be much shorter than the period of the highest frequency wave component under consideration. In our transmission line metamaterial, the duration of the time-interface is primarily dictated by the rise and fall times of the switches. Hence, we experimentally characterize the switching speed of our circuit, and numerically investigate its implications on our implemented time-interface.

To characterize the circuit switching speed, we fabricate a single unit cell of the TLM. We feed a 100MHz sinusoidal signal into the input port, and examine the envelope of the transmitted waveform. Changing the state of the switch with a control signal will result in a transition in the output envelope, whose rise/fall time can be used to quantify the "sharpness" of our time-interface. In this work, we define the interface duration as the time it takes for the output envelope to make a 10%-to-90% or a 90%-to-10% transition. In our measurement, the interface duration is measured to be approximately 3ns, which agrees with the specifications supplied by the manufacturer of the switches.

We also utilize different forms of control signal such as a square wave or a triangular wave. As seen from the measured output envelope in Fig. S10 (black curves), the slew rate of the control signal has no appreciable influence on the duration of the time-interface. As long as $V_{sw}$ can pass the required activation threshold, the switches will have the same transient response. This is an important property since in a realistic operating scenario, the slew rate of the control signal will be limited by the input reactance of the switches. For example, Fig. S11 depicts the actual form of the control signal ($V_{TTL}$) used in obtaining Fig. 2e of the main text.

The slow rise and fall times of the switch input voltage also makes it difficult to precisely define the duration of a temporal slab. Hence, in the main text, we use $\tau$ to refer to the "ON"-state duration of the control signal supplied by the function generator, which is a well-defined rectangular pulse train.

## S10. Effect of Finite Switching Time

In this section, we numerically investigate the influence of finite circuit switching time on the scattering properties of our time-interfaces. First, we compare the time reflections generated by an ideal time-interface (instantaneous switching) and a realistic interface with 3ns 10%-to-90% switch rise time (which corresponds to our experiment). We first inject an asymmetric pulse into the circuit, and probe the time-reflected and time-refracted waveforms at the input and output ports in the time domain. The results are plotted in Fig. S12a. Here, the black curve corresponds to the simulated voltages when the switches have almost instantaneous rise times, while the red curves correspond to the voltages when the rise time is 3ns. It can be seen that the TR and TT waves are almost the same for both cases, which suggests that our platform behaves almost the same as a true time-interface for the frequency components contained in the asymmetric pulse.

In Fig. S12b, we plot the peak amplitude of the time reflection as a function of the rise time, normalized against the peak value when $\tau_{rise} \approx 0$. It can be seen that the 3ns 10%-to-90% rise time produced a time reflection with amplitude that is approximately 90% of the ideal output. The match would be even better when operating at lower frequencies.

## S11. Measurement of Inverted Temporal Slab

In this section, we report measurements of inverted temporal slabs formed by an "ON-OFF-ON" switching sequence, with different "OFF"-durations $\tau$. This provides the complementary control experiment for the "OFF-ON-OFF" temporal slabs discussed in the main text (Fig. 3c, 3d and 3e). It should be reiterated that $\tau$ corresponds to the "OFF" duration of the control signal (an ideal rectangular pulse train). It does not equal to the actual duration of the realized temporal slab, due to the input reactance of the switches. However, as discussed in the section "Characterization of the Circuit Switching Speed", this does not degrade the sharpness of the "facets" of our temporal slab.

In the experiment, we launch broadband pulses into the TLM, and observe the TR output produced by the inverted temporal slab. The time domain measurements for three different values of $\tau$ are plotted in Fig. S13a. In each case, two TR pulses with inverted polarity compared to the input can be observed. In contrast with the "OFF-ON-OFF" temporal slab examined in the main text, which produced two TR pulses with almost equal magnitudes, here we see that the second pulse is much more attenuated than the first. The amplitude difference is exacerbated as $\tau$ is increased. This can be explained by the loss of the TLM in its "ON" state. As we increase $\tau$, the second pulse will have travelled further in the TLM before the ending of the temporal slab. Therefore, it must traverse a longer distance inside the lossy TL before returning to port 1, meaning it will experience more attenuation.

We carry out the same Fourier analysis as is done in the main text, to obtain the total reflection spectrum of each inverted temporal slab (Fig. S13b). As expected, we observe reflection minima for certain frequencies. The location of the dips can be tuned by adjusting $\tau$. Due to the strong dissipative loss in the "ON"-state of the TLM, we see that the reflections never reach exactly zero. This is reminiscent of the properties a lossy Fabry-Perot etalon.

## S12. Video Leakage of the Switches

Video leakage refers to the generation of spurious signals at the RF ports of an integrated circuit switch, typically caused by the switch driver when a large voltage spike (such as that caused by our control signal) occurs at the logic input. Such signals can travel through our transmission line and contaminate the voltage measurements at the input and output ports, especially when it overlaps with our signal pulses and/or their temporal reflections, as seen by the upper plot in Fig. S14.

Although video leakage is unavoidable during measurement, we can remove its effect through post processing. We use a logic control signal with higher frequency than the signal pulse train. This will produce multiple identical copies of the video leakage signal, with period $T$. Some copies will inevitably overlap with the signal pulses (which have period $T_s$). However, most of the copies of the video leakage will appear in between the signal pulses. Since the leakage is almost identical during each switching cycle, we can simply subtract the measured waveforms by its copy, shifted by $T$. This will remove all video leakage, while leaving intact the desired signal.

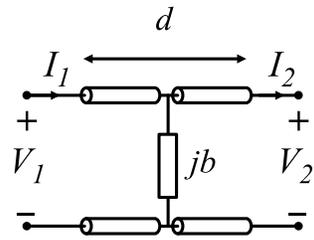

**Fig. S1.**

Equivalent T-network for one unit cell of our TLM when the switches are closed. Here, $d$ is the physical length of the unit cell (0.2080m) and $jb = j2.753$ is the normalized load susceptance at 100 MHz.

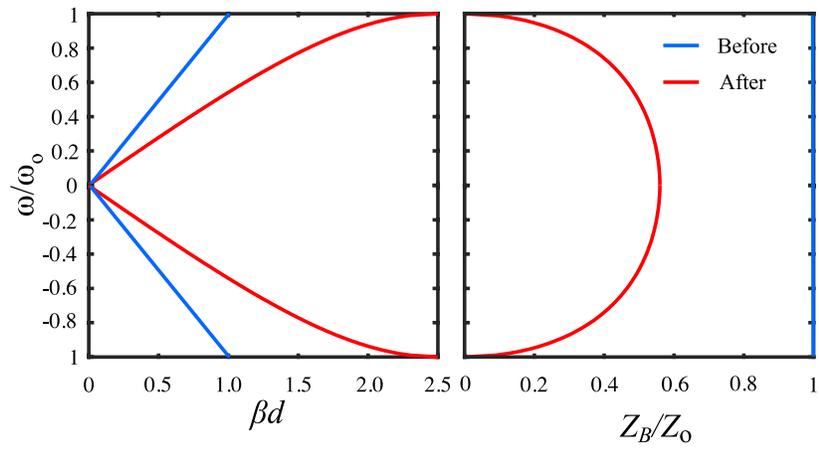

**Fig. S2.**
Normalized dispersion relation and Bloch impedance of our transmission line, assuming an unit-cell electrical length of $\theta = kd = 72°$, and an normalized load susceptance of $b = 2.753$, both evaluated at 100 MHz. The blue lines and the red lines correspond to the unloaded and the loaded lines respectively.

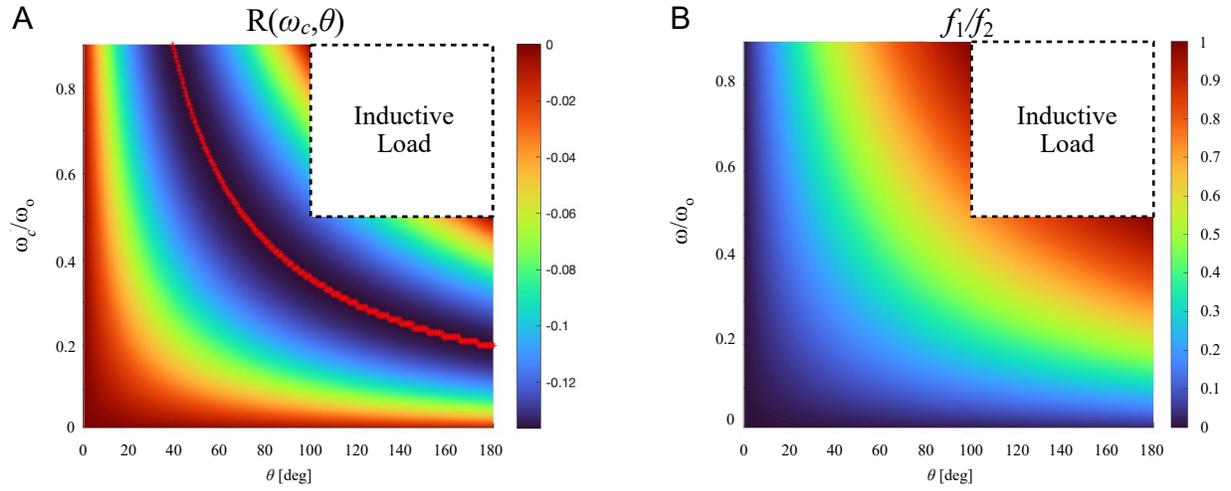

**Fig. S3.**
Parametric study for the identification of the optimal unit-cell electrical length $\theta$ and cutoff frequency of the first band gap $\omega_c$, which is related to the load susceptance $b$. (**A**) A map for the temporal reflection coefficient and (**B**) a map for the frequency translation ratio. The red line in panel A corresponds to the combination of $(\theta, \omega_c)$ that yields the largest reflection coefficient. The white boxes with dotted border in both panels denote regions in which the load susceptance become inductive (i.e., negative).

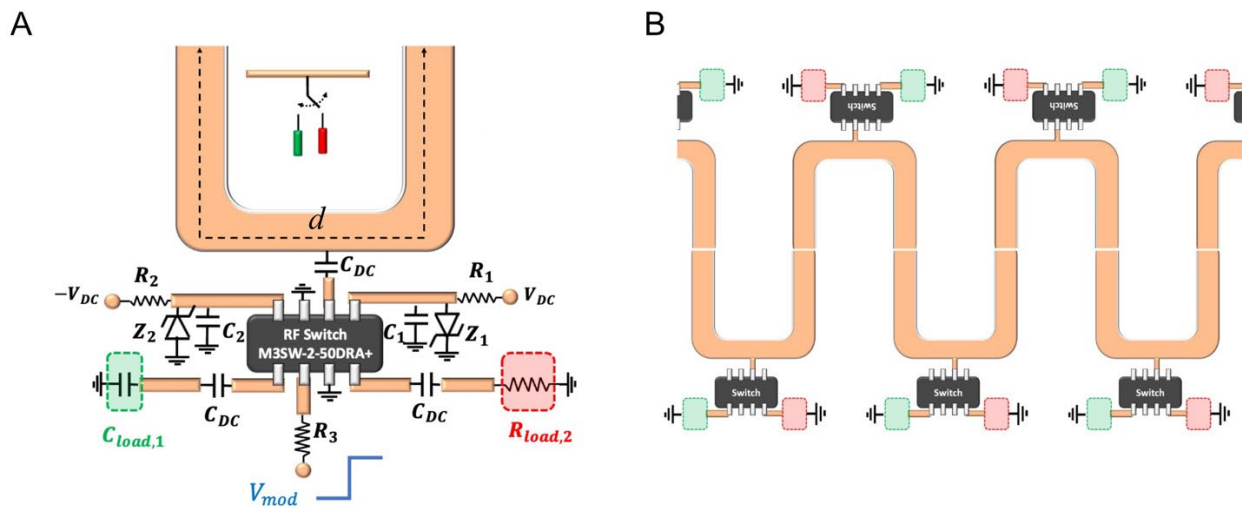

**Fig. S4.**

Circuit layout for the fabricated transmission line. (**A**) Layout for the unit cell. The descriptions of the components can be found in Table **S1**. (**B**) Schematic for the complete TLM, showing how the unit cells are connected.

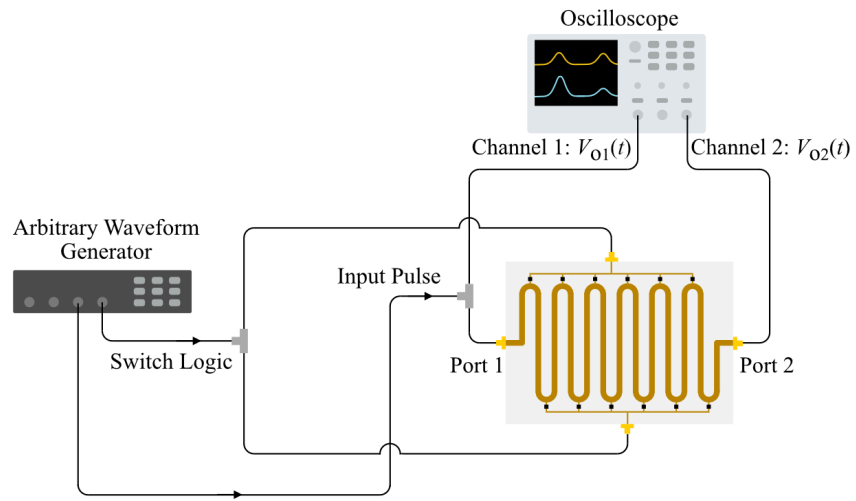

**Fig. S5.**

Simplified schematic of the experimental set up for time domain measurements. Power supply for the switches and a separate function generator for switch control are not shown.

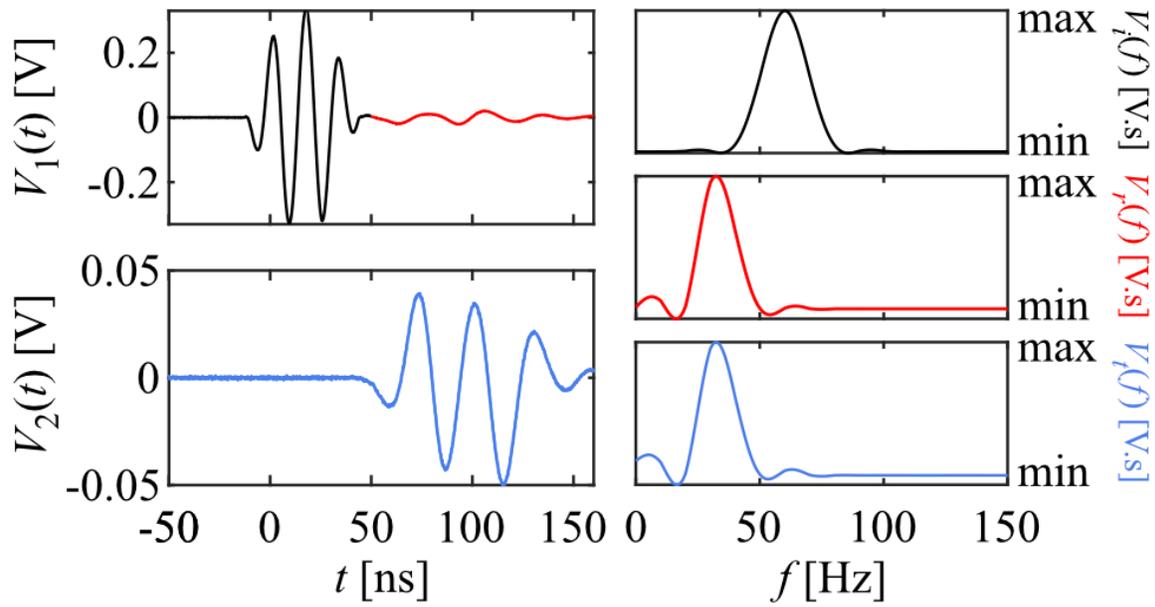

**Fig. S6.**

A sample of the measurement used to characterize the frequency conversion of our time-interface when the wave impedance reduced. The left column shows the incident (black), time-reflected (red) and time-refracted (blue) signals. The right column shows their corresponding Fourier transform. A frequency down conversion by about 0.55 is clearly visible through the peak of the frequency spectra.

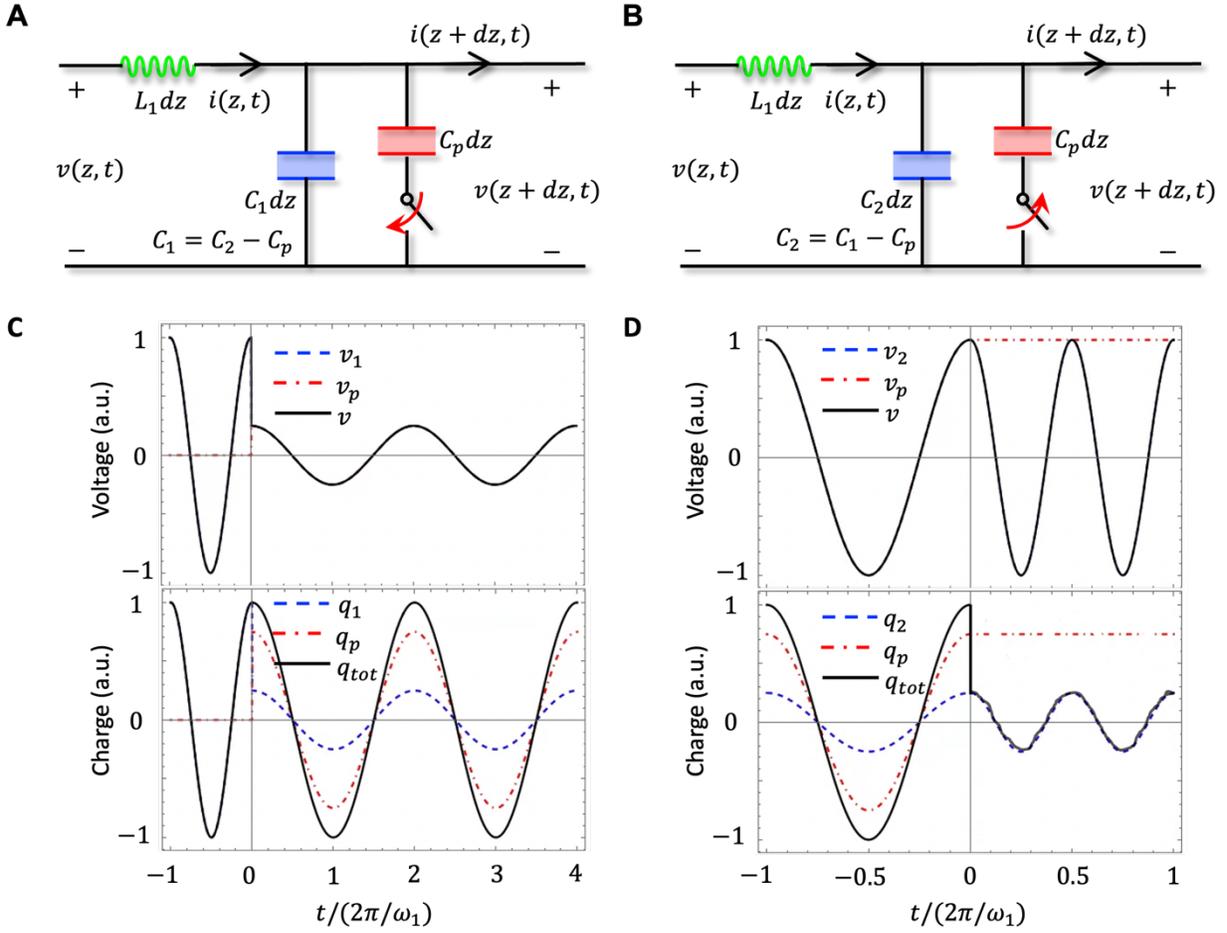

**Fig. S7.**
Circuit model of time-switched transmission-line (TL), with the (**A**) loading, and (**B**) removal of a parallel capacitor. $C_1$ ($C_2$) denotes the value of effective distributed capacitance of the TL before (after) closing or opening the switch. Panel **C** demonstrates the time variation of voltages and charges on each capacitor, as well as their total values sensed by the TL, corresponding to a numerical example of Panel **A** with $C_2 = 4C_1$. Panel **D** shows the same quantities for a numerical example of Panel **B** with $C_2 = C_1/4$.

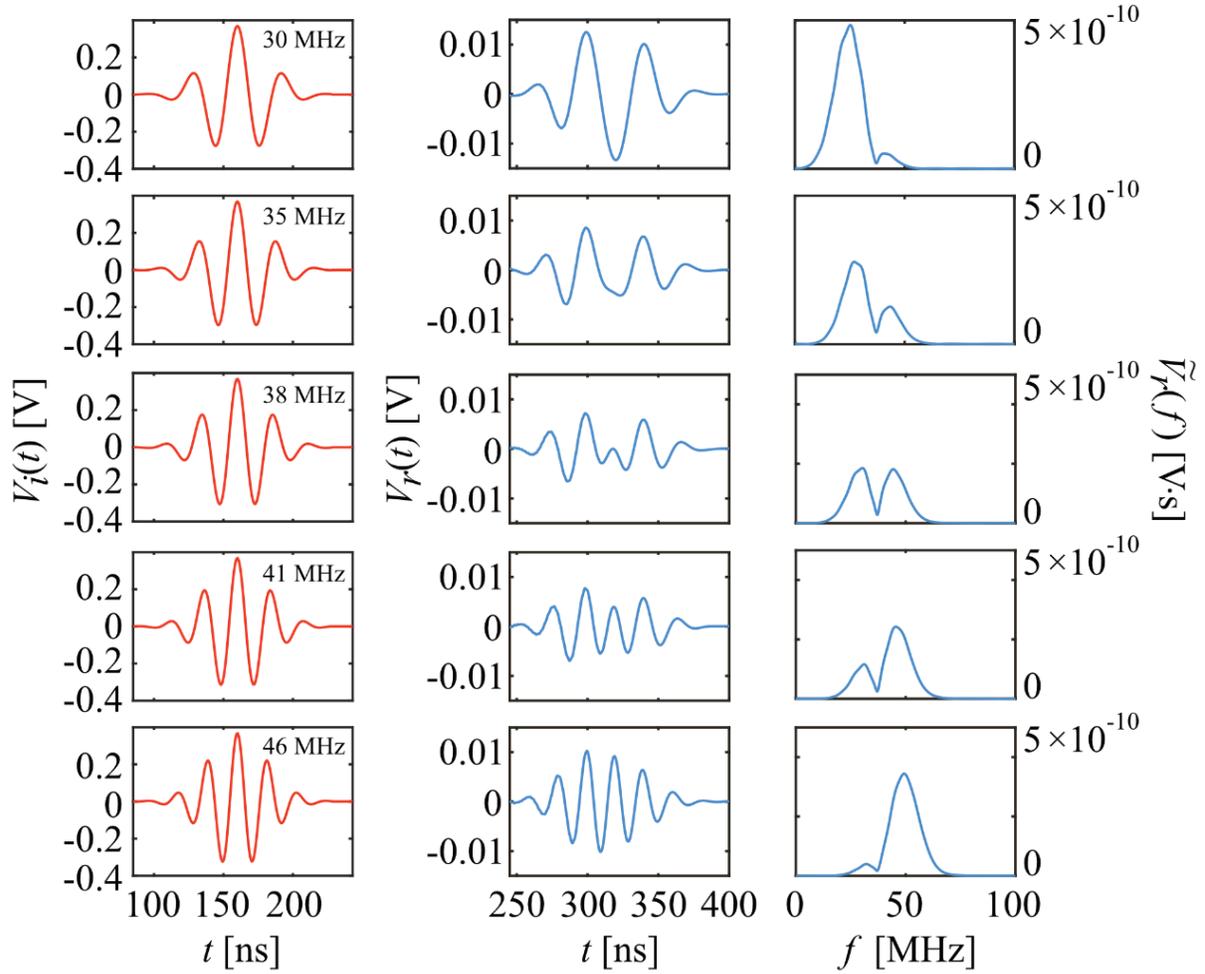

**Fig. S8.**
Circuit simulations for a temporal slab with a fixed "ON" time ($\tau$ =15ns). The injected pulses are in the form of Gaussian wave packets with different carrier frequencies and 17.5MHz spectral FWHM. The simulated circuit consists of 3 cascaded copies of our TLM sample. The left column depicts the time domain waveform of the input pulse, while the middle column shows the total time-reflected output. The right column shows the frequency spectra of the output. All of the spectra exhibit reflection nulls at approximately 38MHz, which agrees with measured results in Fig. 3d of the main text.

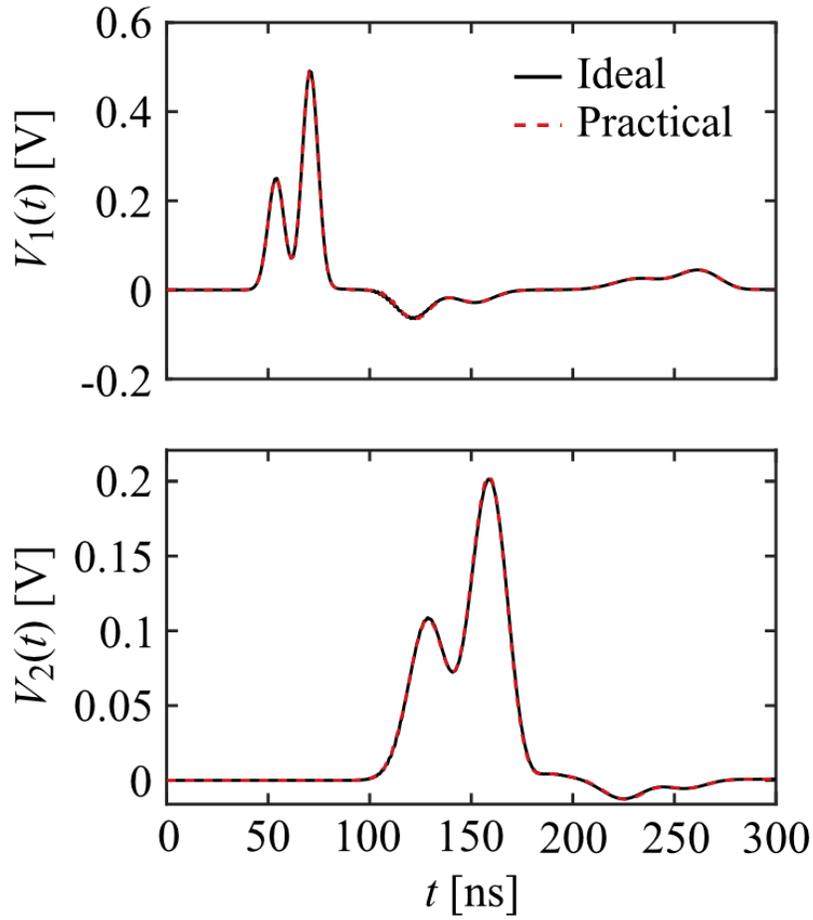

**Fig. S9.**

Circuit simulations demonstrating the effect of switch desynchronization. The black curve corresponds to the simulated voltages at ports 1 and 2 assuming perfect synchronization. The red curve ("Practical"), which matches the ideal one almost exactly, depicts the simulation of our measurement set up. This shows that despite the finite time delay between adjacent unit cells, our set up can emulate a homogeneous time-interface with very high fidelity.

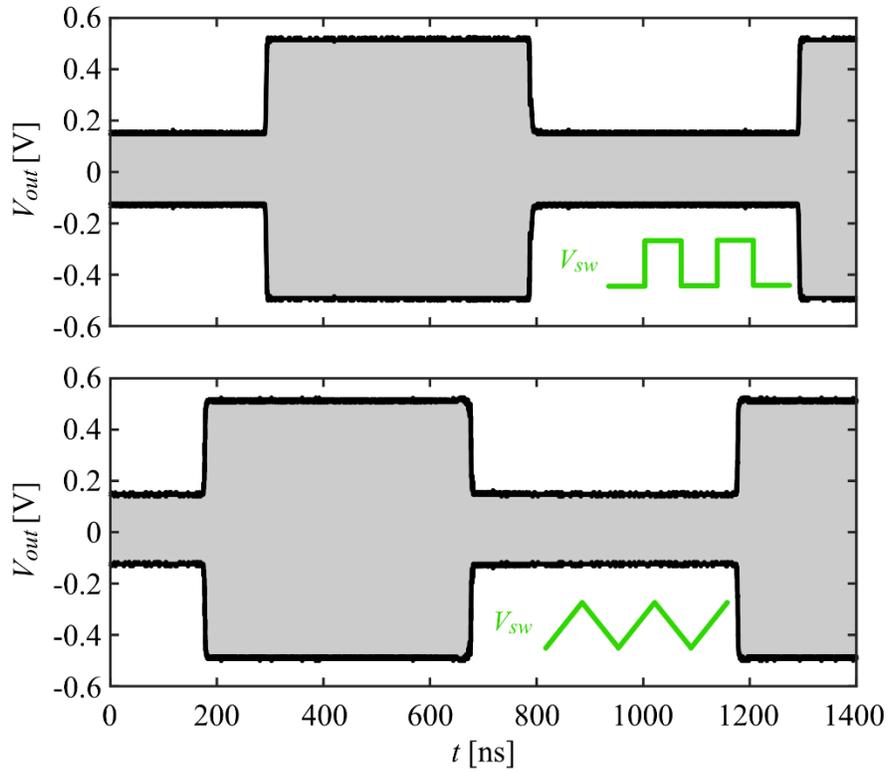

**Fig. S10.**

Experimental characterization of the circuit switching speed using different types of control signal. Here, a 100MHz sinusoidal signal is transmitted through a single unit cell sample of the TLM, and the envelope of the output waveform is recorded (black curve). We activate the switch inside the unit cell using different types of control signals $V_{sw}$, with drastically different slopes. It can be seen that both a rectangular control signal (top) and a triangular control signal (bottom) produced identical output envelopes, demonstrating that the rise time of $V_{sw}$ has very little influence on the switching speed of the circuit.

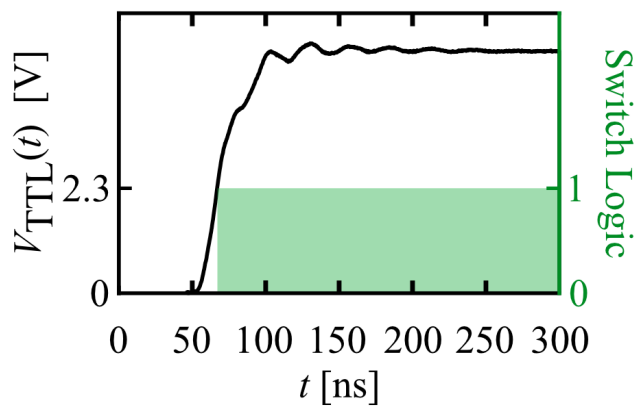

**Fig. S11.**
A sample measurement of the control signal used to induce the time-interface. The slew rate of the signal is very low, due to the input reactance of the switches. However as shown in Fig. S10 that, simply crossing the activation threshold of the switch (2.3V according to the datasheet) is sufficient to induce a sharp time-interface. This particular control signal was used to induce the time-interface examined in Fig 2e of the main text.

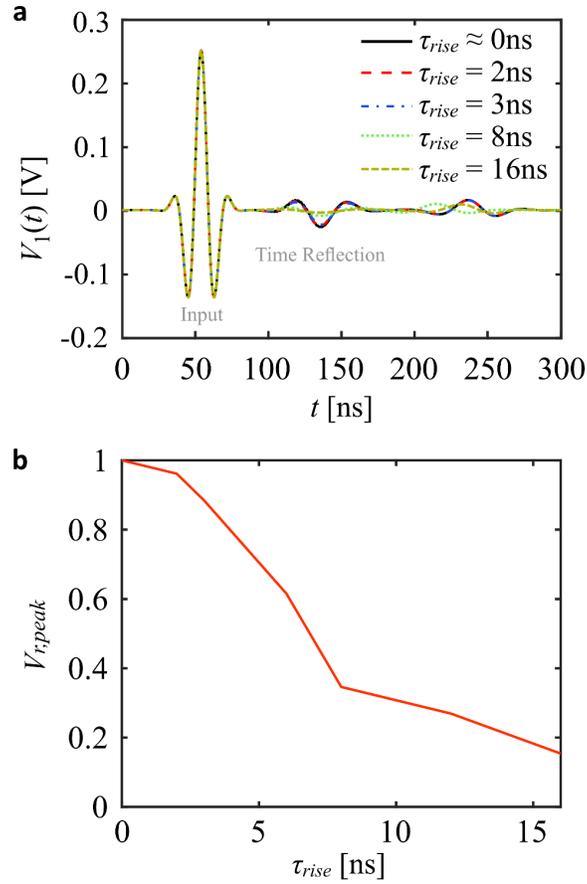

**Fig. S12.**
Circuit simulation of the TLM with different 10%-to-90% rise time for the switches. (a) Time domain voltage measured at the input port. The input is a Gaussian wave packet with 50MHz center frequency and 40MHz spectral FWHM. The blue curve represents the simulation results when the circuit has the same rise time (3ns) as our selected RF switches; it is seen to closely match the black curve, which corresponds to the case when the rise time is near zero. As we further increase the rise time to 8ns (green) and 12ns (yellow), the amplitude of the time reflection become significantly weaker. (b) Peak amplitude (normalized against the ideal case) of the time reflected wave packet as a function of the switch rise time. The amplitude corresponding to the case of 3ns rise time was approximately 90% of the ideal amplitude.

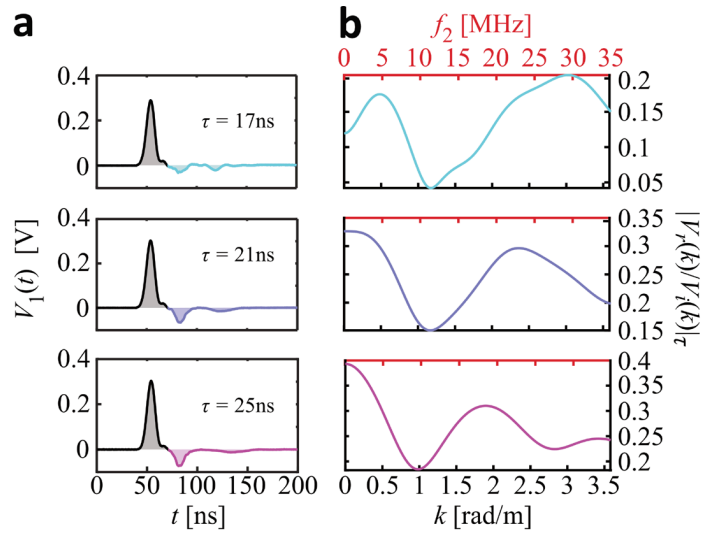

**Fig. S13.**
Measurement of temporal interference induced by the inverted temporal slab. (a) Time domain waveform recorded at port 1 for various values of "OFF" duration $\tau$. As $\tau$ is increased, so does the separation between the two TR pulses. The second TR pulse is heavily attenuated for larger values of $\tau$, since it experiences increased attenuation. (b) Reflection spectra of inverted temporal slab in the wavenumber (bottom axis) and frequency (top axis) domain. We see distinct reflection minima whose locations are tunable via adjustment of $\tau$.

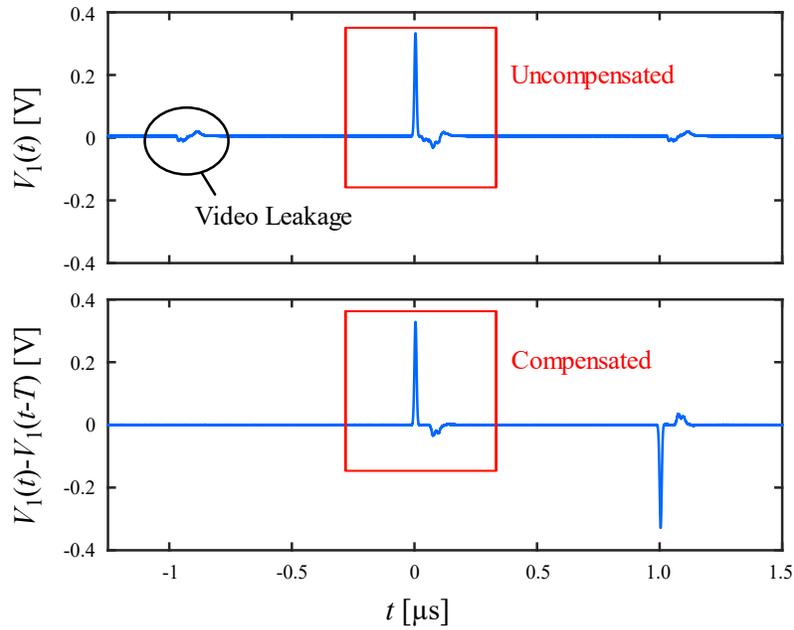

**Fig. S14.**

Compensation of video leakage from the switches. Top plot shows the uncompensated waveform measured at the input port of the TL, with multiple spurious spikes caused by the rising and falling edges of our control signal. The bottom plot shows the compensated waveform, obtained through subtracting the top plot by a shifted copy of itself. The compensated waveform has much more clearly observable input and TR pulses. A spurious inverted copy of the main signal remains, but can be discarded during analysis.

**Table S1.**
List of components used to fabricate our transmission line, referred to the schematic shown in Fig. S4.

| Component | Description | Value | Component Number |
|---|---|---|---|
| $C_{DC}$ | DC blocking capacitor | 3.900 nF | C0402C392K5RACAUTO |
| $C_{1,2}$ | RF shorting capacitor | 10 pF | 06031A100GAT2A |
| $R_{1,2}$ | Resistor | 11.5 Ohm | RK73H1JTTD11R5F |
| $R_3$ | Resistor | 100 Ohm | RK73H1JTTD1000F |
| $Z_{1,2}$ | Zener Diode | $V_z = 5.3\,V$ | MMSZ4690T1G |
| RF Switch | | | M3SW-2-50DRA+ |
| $R_{load,2}$ | TL load (OFF-State) | 1 MOhm | RCS12061M00JNEA |
| $C_{load,1}$ | TL load (ON-State) | 82 pF | C0402C823K9RACTU |

# References

S1. Keysight, PathWave Advanced Design System. *Keysight*, (available at https://www.keysight.com/us/en/products/software/pathwave-design-software/pathwave-advanced-design-system.html).
S2. D. M. Pozar, *Microwave Engineering* (John Wiley & Sons, 2011).
S3. Minicircuits Fast Switching SPDT RF Switch, (available at https://www.minicircuits.com/pdfs/M3SW-2-50DRA+.pdf).
S4. Pacheco-Peña, V. & Engheta, N. Antireflection temporal coatings. *Optica* **7,** 323–331 (2020).
S5. Galiffi, E., Yin, S. & Alú, A. Tapered photonic switching. *Nanophotonics* **11**, 3575–3581 (2022).
S6. Hadad, Y. & Shlivinski, A. Soft Temporal Switching of Transmission Line Parameters: Wave-Field, Energy Balance, and Applications. *IEEE Transactions on Antennas and Propagation* **68**, 1643–1654 (2020).